\documentclass{moriond}

\def\be{\begin{equation}}
\def\ee{\end{equation}}
\def\bea{\begin{eqnarray}}
\def\eea{\end{eqnarray}}

\newcommand{\Photo}{\includegraphics[height=35mm]{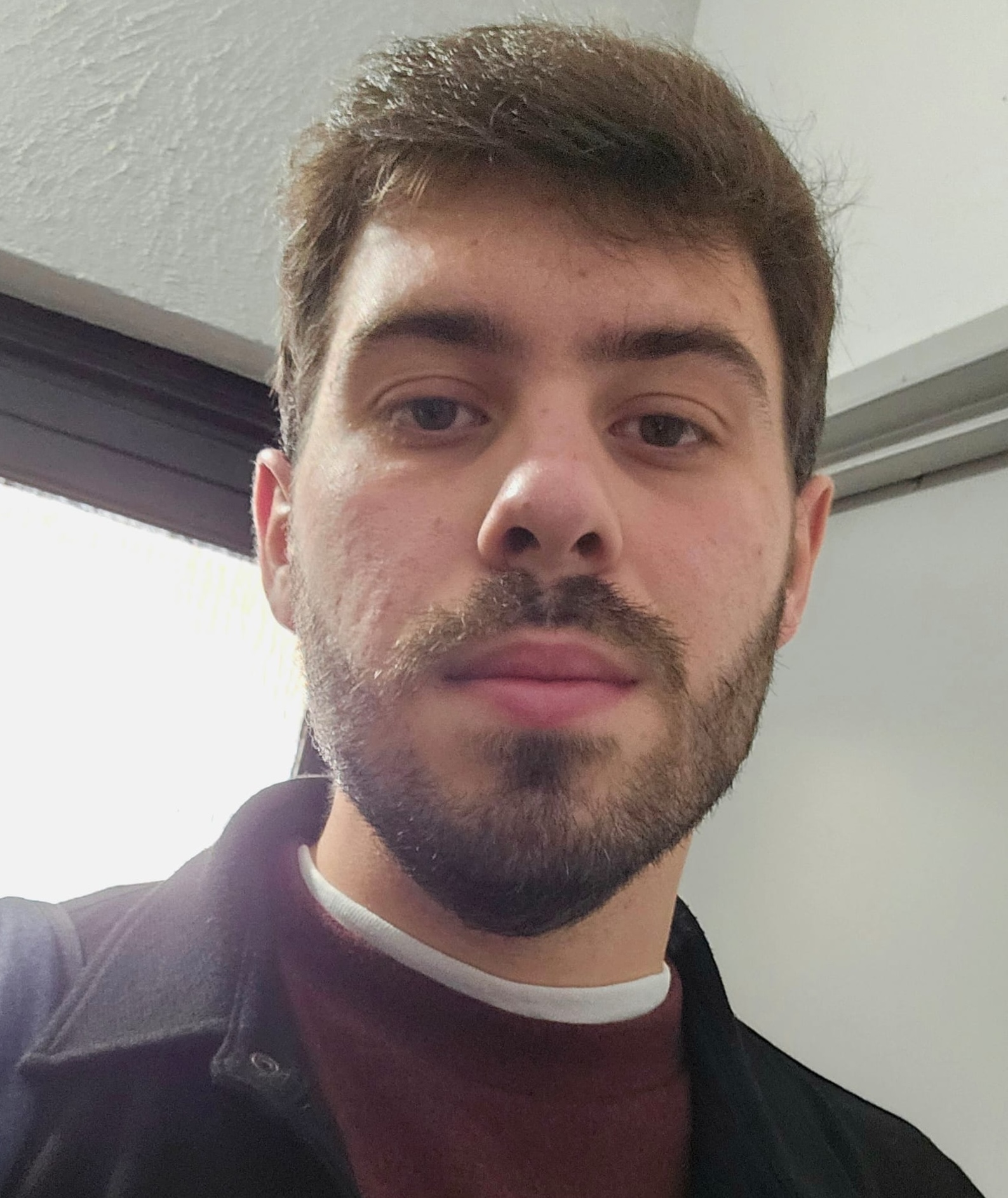}}

\begin{document}
\vspace*{4cm}
\title{CMB lensing imprints of cosmic voids in DESI Legacy Survey DR9 LRGs with photometric redshift calibration}

\author{ Simone Sartori }

\address{INAF-IASF Milano, Via Alfonso Corti 12, 20133 Milano, Italy}

\maketitle\abstracts{
The lensing imprint of cosmic voids in the Cosmic Microwave Background (CMB) provides a powerful test of the $\Lambda$CDM model. However, recent studies report a "lensing-is-low" tension between observations and mock predictions. To investigate this, we measure the stacked CMB lensing signal of 3D cosmic voids identified in the DESI Legacy Survey DR9 Luminous Red Galaxy (LRG) sample, cross-correlated with the Planck 2018 lensing map. We compare our observations to $\Lambda$CDM templates derived from Buzzard mocks, critically calibrated using over one million DESI spectra to perfectly match the sparseness and photometric redshift error distributions of the observed data. By categorizing voids based on their gravitational potential, we disentangle the signals of void-in-voids and void-in-clouds, achieving two independent record detection significances of $\sim 17\sigma$. We find full agreement between observations and simulated $\Lambda$CDM templates across all void populations and redshift bins ($0.35\! <\! z\! <\! 0.95$), measuring an amplitude parameter $A_\kappa = 1.016 \pm 0.054$ for the full sample. This highlights the necessity of accurate systematic control, effectively resolving the reported lensing tension within this dataset. This proceeding summarizes the results presented in Sartori et al. (2025)\cite{sartori2025}.}

\section{Introduction}

The lensing imprint of cosmic voids in the Cosmic Microwave Background (CMB) provides a unique probe to test the $\Lambda$CDM cosmological model. Voids induce a demagnification effect on the CMB, resulting in a negative signal in the lensing convergence ($\kappa$) maps. Because the signal from an individual void is entirely dominated by instrumental and background noise, standard analyses rely on stacking techniques over large catalogs. While this methodology has successfully yielded detections with high significance \cite{Cai2014,Raghunathan2020}, several recent studies have highlighted a puzzling ``lensing-is-low'' tension, reaching up to $\sim\! 3\sigma$, between the observed stacked signals and $\Lambda$CDM simulated predictions \cite{Vielzeuf2021,Hang2021,Kovacs2022,CamachoCiurana2024}. The amplitude of this discrepancy is highly dependent on factors such as void identification strategies and the specific properties of the tracer galaxy sample. Consequently, uncontrolled mismatches between observations and mocks could act as severe hidden systematics, mimicking a breakdown of the standard model.

To address this tension, we exploit the unprecedentedly vast Luminous Red Galaxy (LRG) dataset from the DESI Legacy Survey DR9 \cite{Dey2019,Zhou2023}, comprising approximately 10 million LRGs distributed over $\sim\! 19,500$ deg$^2$ in the redshift range $0.35\! <\! z\! <\! 0.95$, and cross-correlate them with the Planck 2018 CMB lensing convergence map \cite{Planck2020}. Crucially, we compare our observations with $\Lambda$CDM templates derived from the DESI-like Buzzard mocks \cite{DeRose2019}, which we explicitly calibrated using over one million DESI spectra to correct and model their photometric redshift errors. By carefully matching the sparseness and photometric redshift error distributions between the observed and simulated LRGs, we ensure a reliable and consistent void identification, effectively suppressing the systematics that could otherwise induce false cosmological tensions.

\section{Data sets}

We identify our cosmic void sample using the LRG population extracted from the ninth Data Release (DR9) of the DESI Legacy Imaging Surveys~\cite{Dey2019,Zhou2023}. The catalog covers a total sky area of 19,573~deg$^2$, combining observations from the MzLS, BASS, DECaLS, and DES surveys. This vast coverage provides a sample of approximately 10.4 million LRGs in the photometric redshift range $0.30\!<\!z_\mathrm{phot}\!<\!1.05$. This data set is supplemented with around one million spectroscopic redshifts from the early stages of the DESI survey for redshift calibration.

For the CMB lensing observable, we utilize the publicly available reconstructed convergence ($\kappa$) map from the Planck 2018 data release~\cite{Planck2020}. The map is reconstructed using the minimum-variance quadratic estimator. To suppress noise at small scales while preserving the imprint of smaller structures, we apply a Gaussian smoothing with FWHM = $0.5^\circ$.

Finally, the theoretical $\Lambda$CDM predictions are derived from four different realizations of the Buzzard mocks~\cite{DeRose2019}. These are quarter-sky simulated catalogs based on $N$-body lightcone simulations initialized with a flat $\Lambda$CDM cosmology ($\Omega_\mathrm{m}=0.286$, $\sigma_8=0.82$, $h=0.7$). The mocks are populated with synthetic galaxies using the \texttt{Addgals} algorithm~\cite{Wechsler2022}, which assigns realistic positions, velocities, and spectral energy distributions to dark matter particles to reproduce the clustering properties of subhalo abundance matching models.

\section{Methodology}

\subsection{Mock Calibration}
To ensure structural consistency between observed and simulated tracer populations, we calibrated the Buzzard mocks following a two-step procedure. Because mismatches in galaxy properties can propagate into voids and bias the lensing signal, we first corrected the mock photometric redshift error distributions by sampling the $z_\mathrm{spec}-z_\mathrm{phot}$ distribution derived from over one million DESI spectra. Subsequently, we performed a redshift-dependent \textit{sparseness matching} by downsampling the mocks to align their number density with the observed LRGs in $\Delta z = 0.01$ bins. This rigorous injection of observational systematics ensures that the simulated voids are subject to the same selection effects and density profiles as the observed ones.

\subsection{Void Finding and Classification}
We identified 3D cosmic voids using the \texttt{REVOLVER} algorithm~\cite{Nadathur2019}, which utilizes a Voronoi tessellation of the tracer density field to define underdense basins associated to local density minima. Restricting the void centers to the redshift range $0.35\! <\! z \!<\! 0.95$ yielded a full-sky catalog of 140,712 observed voids, constituting the largest sample of its kind to date. 

To disentangle the distinct lensing imprints produced by different large-scale environments, we categorized voids using the $\lambda_\mathrm{v}$ parameter~\cite{Nadathur2017}, defined as $\lambda_\mathrm{v} = \bar{\delta}_\mathrm{v} [R_\mathrm{v} / (1 h^{-1}\mathrm{Mpc})]^{1/2}$, where $\bar{\delta}_\mathrm{v}$ is the mean density contrast and $R_\mathrm{v}$ is the effective void radius. This parameter serves as a proxy for the central gravitational potential, allowing us to define three populations: highly underdense \textit{void-in-voids} ($\lambda_\mathrm{v}\! <\! -5$), a transition sample ($-5\! \le\! \lambda_\mathrm{v}\! <\! 5$), and slightly underdense \textit{void-in-clouds} ($\lambda_\mathrm{v}\! \ge\! 5$). Furthermore, we divided the catalogs into four equally spaced redshift bins of $\Delta z = 0.15$ to perform a tomographic analysis of the signal.

\subsection{Cross-correlation and Error Estimation}
For each void, we extracted a CMB convergence patch of side length $5R_\mathrm{v}$. The stacked radial profile was measured using 25 radial bins, accounting for the local mean convergence bias. The covariance matrix $C_{ij}$ was estimated using 1,000 random CMB maps generated from the Planck power spectrum. We quantified the consistency between observations and simulated templates by fitting the amplitude parameter $A_\kappa$ through a standard $\chi^2$ minimization, incorporating the Anderson-Hartlap correction factor \cite{Hartlap2007} in the inverse covariance matrices.

\section{Results and Discussion}

Following the procedures outlined in the previous sections, we measure the cross-correlation between the cosmic voids identified in the DESI Legacy Survey DR9 LRGs and the Planck 2018 CMB lensing map. We further estimate the deviation from the $\Lambda\mathrm{CDM}$ predictions by comparing the observed signals with the spectroscopically calibrated templates from the Buzzard mocks. 

\begin{figure}[h]
\centering
\includegraphics[width=0.8\textwidth]{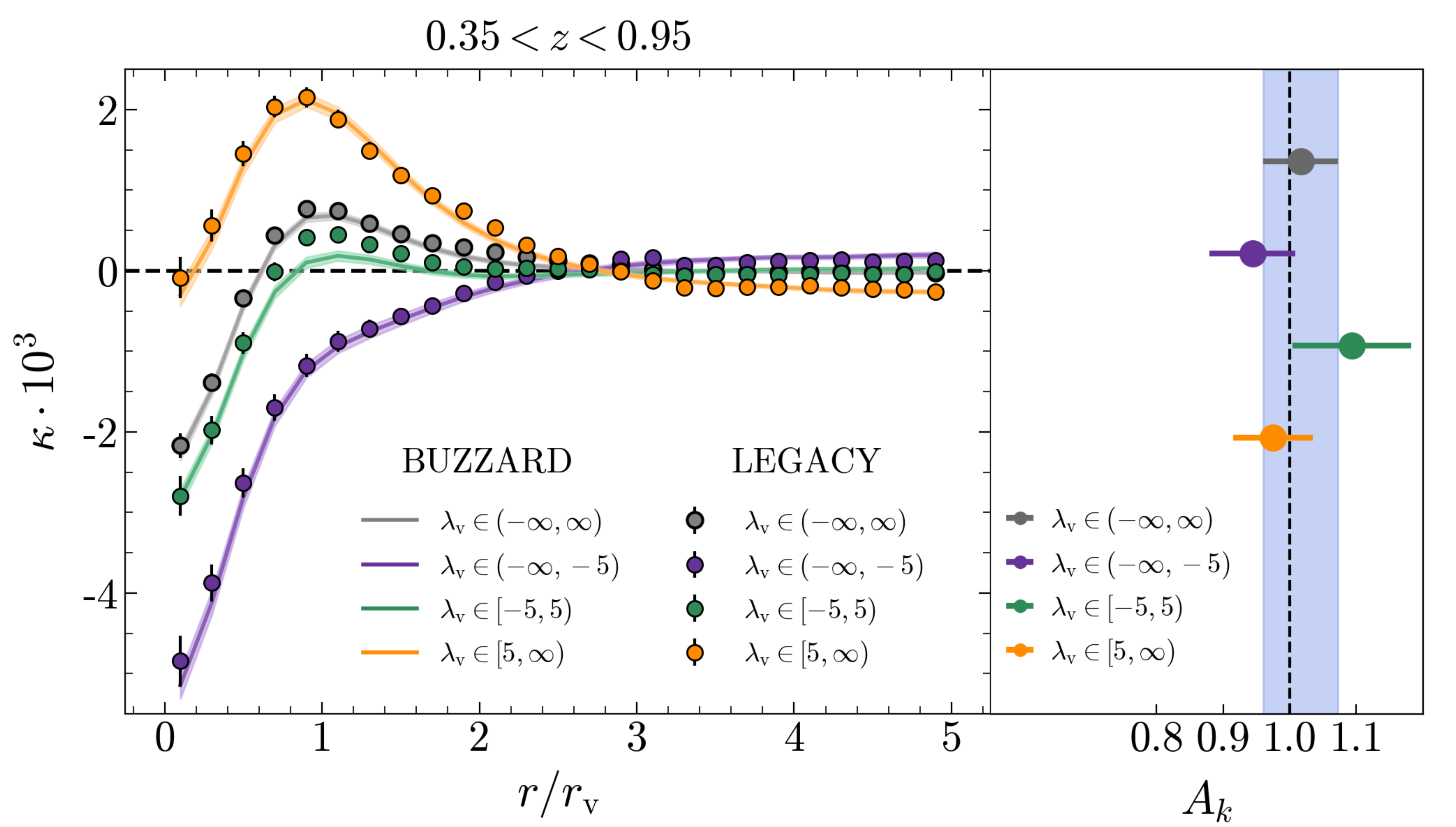}
\caption{Left: Cross-correlation signals for voids identified in the full-sky DESI Legacy Survey (dots) and corresponding calibrated Buzzard mock realizations (lines). Measurements are provided for the full void samples and the three different $\lambda_{\mathrm{v}}$ bins considered. Right: Summary of $A_{\mathrm{\kappa}}$ values for the various measurements. The blue vertical band indicates the $1\sigma$ confidence intervals for the full void samples.}
\label{fig:stacking}
\end{figure}

The results of this full-sky analysis are summarized in Fig. \ref{fig:stacking}. We measure a lensing amplitude of $A_{\mathrm{\kappa}} = 1.016 \pm 0.054$ with a signal-to-noise ratio ($S/N$) of 14.06 for the entire void sample. By categorizing the voids into three distinct populations based on their $\lambda_{\mathrm{v}}$ value, we successfully disentangle their unique physical signatures. For \textit{void-in-voids} and \textit{void-in-clouds}, we find $A_{\mathrm{\kappa}} = 0.944 \pm 0.064$ ($S/N = 16.94$) and $A_{\mathrm{\kappa}} = 0.975 \pm 0.060$ ($S/N = 17.02$), respectively. These results represent the highest detection significances to date for studies of this kind.

\begin{figure}[h]
\centering
\includegraphics[width=0.97\textwidth]{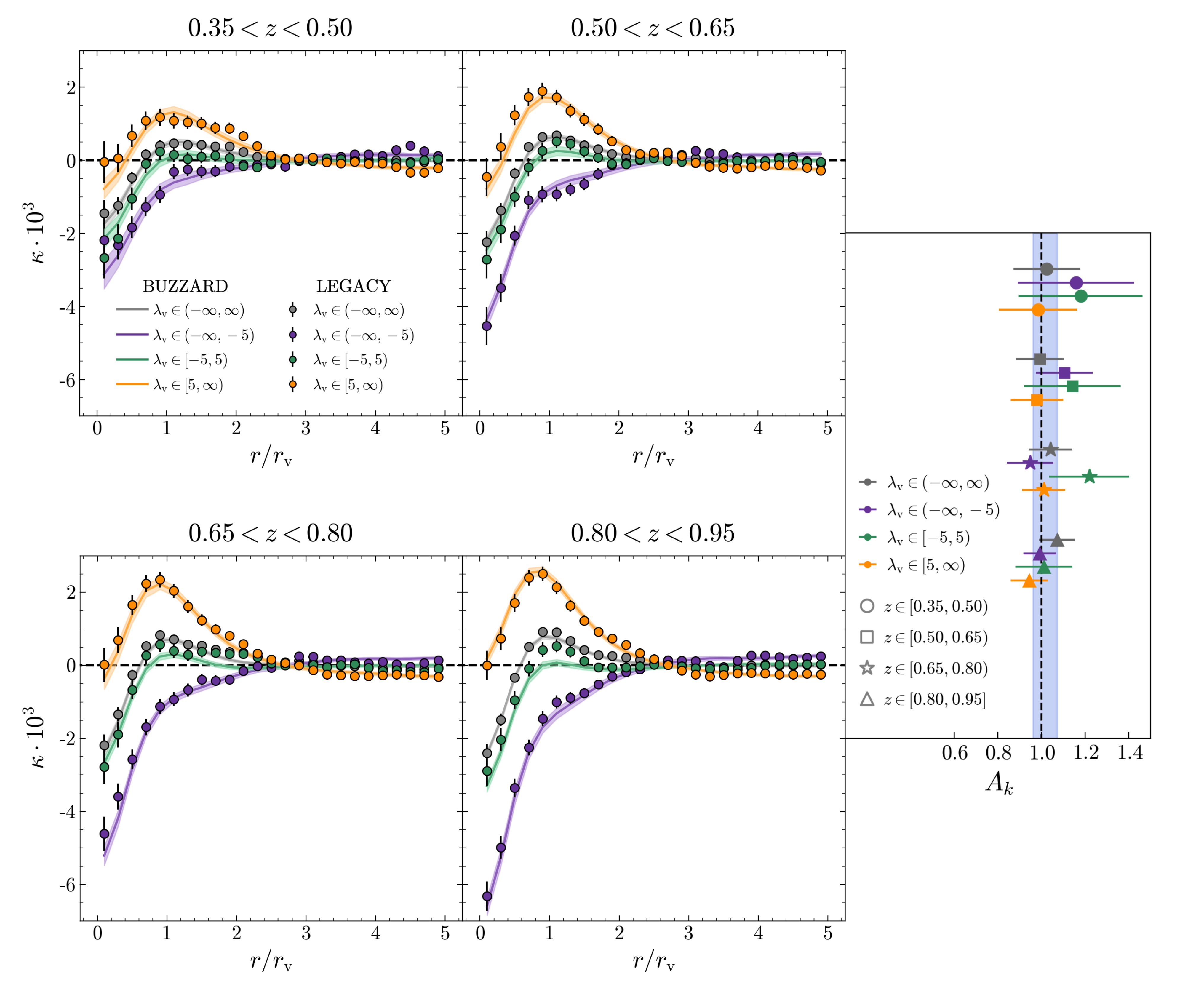}
\caption{Left: Tomographic cross-correlation signals in four equispaced redshift bins ($\Delta z=0.15$). Right: Summary of $A_{\mathrm{\kappa}}$ values. Symbols and void populations follow the same conventions as in Fig.~\ref{fig:stacking}.}
\label{fig:tomography}
\end{figure}

To validate these findings across cosmic time, we perform a tomographic study by dividing the redshift range into four equally spaced bins with $\Delta z = 0.15$. The results, presented in Fig.~\ref{fig:tomography}, show that all observed cross-correlations are consistent with the mock-derived $\Lambda\mathrm{CDM}$ predictions at the $\sim 1\sigma$ level. The amplitude parameter remains remarkably stable and close to unity ($A_{\mathrm{\kappa}} \sim 1$) throughout the entire redshift range ($0.35\! < \!z\! <\! 0.95$), confirming that the temporal evolution of the lensing imprint follows standard theoretical expectations.

The complete absence of the ``lensing-is-low'' tension in our analysis highlights the critical role of systematic management. Our findings suggest that previously reported discrepancies were primarily driven by mismatches in sparseness and photometric redshift error distributions between mocks and observations. By accurately matching these properties using over one million DESI spectra, we effectively suppress these tensions. Ultimately, this work demonstrates that discriminating void populations is essential to significantly boost the S/N and reveal the redshift evolution of their distinct physical signatures.

\section*{References}
\bibliography{moriond}


\end{document}